\documentclass[reprint,prd,aps,nofootinbib,longbibliography]{revtex4}
\usepackage[utf8]{inputenc}
\usepackage{xcolor}
\usepackage{verbatim}
\usepackage{amsmath}
\usepackage{amssymb}
\usepackage{cancel}
\usepackage[bookmarks=false,
 breaklinks=false,pdfborder={0 0 1},backref=section,colorlinks=false]
 {hyperref}
\hypersetup{
 bookmarks}

\makeatletter
\@ifundefined{textcolor}{}
{%
 \definecolor{BLACK}{gray}{0}
 \definecolor{WHITE}{gray}{1}
 \definecolor{RED}{rgb}{1,0,0}
 \definecolor{GREEN}{rgb}{0,1,0}
 \definecolor{BLUE}{rgb}{0,0,1}
 \definecolor{CYAN}{cmyk}{1,0,0,0}
 \definecolor{MAGENTA}{cmyk}{0,1,0,0}
 \definecolor{YELLOW}{cmyk}{0,0,1,0}
}

\usepackage{epsfig}\usepackage{mathrsfs}

\usepackage{subfigure}
\usepackage{times}
\usepackage{yhmath}
\usepackage[normalem]{ulem}
\usepackage{soul}

\newcommand{\lms}{\ell_{{\text ms}}}
\newcommand{\lmfp}{\ell_{\text{mfp}}}

\newcommand{\dvol}{\mbox{dvol}}
\newcommand{\bessel}{\textrm{\textbf{K}}}

\newcommand{\p}{\textrm{p}}

\numberwithin{equation}{section}

\makeatother

\begin{document}
\title{First-order general constitutive equations for relativistic fluids
using the projection method in the Chapman-Enskog expansion of the
Boltzmann equation}
\author{A. L. García-Perciante$^{1}$, A. R. Méndez$^{1}$, and O. Sarbach$^{2}$}
\affiliation{$^{1}$Departamento de Matemáticas Aplicadas y Sistemas, Universidad
Autónoma Metropolitana-Cuajimalpa (05348) Cuajimalpa de Morelos, Ciudad
de México, México.}
\affiliation{$^{2}$Instituto de Física y Matemáticas, Universidad Michoacana de
San Nicolás de Hidalgo, Edificio C-3, Ciudad Universitaria, 58040
Morelia, Michoacán, México.}

\maketitle

\section*{Abstract}

The first-order out of equilibrium correction to the distribution function, obtained by implementing the projection method for the perturbed relativistic Boltzmann equation using the Chapman-Enskog method, is generalized in order to explicitly include the freedom of choice for frame and representation. It is shown how this procedure leads to general constitutive equations that couple the dissipative fluxes to all derivatives of the state variables (forces), including a weak external electromagnetic field. Special cases of the resulting force-flux relations have been shown to lead to physically sound theories for relativistic fluids.

\section{Introduction}

The Chapman-Enskog (CE) expansion has been widely implemented in various contexts as a successful perturbative solution to the Boltzmann equation. Moreover, the procedure has been rigorously analyzed and proven to lead to a well rounded theory for dissipative hydrodynamics, based on the kinetic theory of gases~\cite{cercigniani1,LaureLectureNotes}.
However, its formal generalization to the relativistic scenario was somehow paused for decades due to the belief that it unequivocally led to unstable first-order theories, similar to the ones proposed by Eckart and Landau \cite{cE1940,LandauLifshitz-Book6}\footnote{See for example Refs.~\cite{wHlL83} and \cite{wHlL85} for proofs regarding generic instabilities in first-order dissipative fluid theories and \cite{PercianteRubioReula} for the stability analysis of the traditional CE solution.}. Such task has only been recently resumed in view of the new proposals of first-order theories for dissipative relativistic fluids which can be shown to lead to stable, hyperbolic, and causal systems of transport equations~\cite{pKovtun19,Bemfica2022,aGjSoS2024a,aGjSoS2024b}.

In recent works, using a rigorous CE implementation that generalizes Saint-Raymond’s formalism~\cite{LaurePaper,LaureLectureNotes} to the relativistic case, we obtained a family of first-order non-equilibrium distribution functions giving rise to constitutive equations featuring generalized forces which contain both spatial and temporal projections of the gradients of the state variables~\cite{JNET24,Hydrolimit}. Moreover, in the presence of external fields this approach can in principle lead to dissipative contributions arising from them. Based on this novel and more formal procedure within the CE hypothesis for the relativistic Boltzmann equation, in Ref.~\cite{Hydrolimit} a relativistic hydrodynamic theory for dissipative fluids was formulated in a particular frame which fixes the particle current and the trace of the energy-momentum-stress tensor to their equilibrium values. The motivation for choosing these conditions is based on the fact that they lead to the solution to the Boltzmann equation to first order in Knudsen's parameter featuring no contribution from the homogeneous solution. That is, one demands orthogonality of the non-equilibrium correction with the kernel of the linearized collision operator. The resulting theory was analyzed and shown to be hyperbolic, causal, and stable in Refs.~\cite{aGjSoS2024a,aGjSoS2024b}.

The particular choice described above, which we refer to as the trace-fixed particle (TFP) frame, is justified by the freedom of choosing the variables describing non-equilibrium states. Moreover, in recent proposals Bemfica, Disconzi, Noronha, Kovtun, and collaborators (see for example~\cite{pKovtun19,Bemfica2022}) have established general first-order theories for relativistic dissipative fluids (known as BDNK theories) and shown that for some other choices of frames the transport equations constitute a well-posed Cauchy problem with causal propagation and stable global equilibria. In those works the analysis is carried out in arbitrary frames, and conditions on the parameters involved for the theory to have the desirable properties are established. Also, in Refs.~\cite{PhysRevD.104.096016,rHpK22,Rocha2022,RDNR2024} the microscopic foundations of these theories are addressed based on different approaches, including the method of moments, Hilbert's method and the CE approximation. However, it is important to point out that the analysis carried out in Ref.~\cite{Rocha2022} for the CE method involves a procedure which eliminates the time derivatives from the first-order solution. Also a modified CE method is proposed in Refs.~\cite{Rocha2022,RDNR2024}, which differs from the approach presented in Refs.~\cite{JNET24,Hydrolimit} and studied in the present work.

The choice of frame in the phenomenological context is imposed by selecting \textit{matching conditions} which are introduced in the general constitutive equations. In this work we discuss how this corresponds, in the microscopic framework, to choosing the compatibility conditions which are required for the uniqueness of the CE solution to the relativistic Boltzmann equation. Moreover, we show how the constitutive equations in other frames can be obtained from the particular solution in the TFP frame by changing the compatibility conditions.  

In addition to the frame choice, there is a second, independent freedom, which we refer to as the choice of representation~\cite{JNET24}. In the phenomenological framework this freedom consists in adding terms to the constitutive equations which are second order on shell. From the microscopic point of view such terms arise naturally by suitable modifications of the CE method by utilizing the balance equations as shown in Ref.~\cite{Hydrolimit}. In this article we show that exploiting both the frame and representation freedoms finally leads to general first-order constitutive equations which are similar to those obtained in BDNK theories. It is important to emphasize that this second freedom is crucial in order to obtain all the force-flux couplings involved in the general theories, since the particular solution to the Boltzmann equation only features certain combinations of gradients of the state variables. More importantly, this step opens the possibility of having a hyperbolic, stable and causal system of transport equations. To the authors' knowledge, the explicit introduction of this representation freedom within the kinetic formalism, which was included in Ref.~\cite{Hydrolimit} for the TFP frame, has not been tackled and thoroughly discussed so far in a rigorous manner. 

Thus, the main purpose of this work is to derive the most general first-order constitutive equations from the CE projection method discussed in Ref.~\cite{Hydrolimit}, by considering an arbitrary frame and representation. Moreover, we also discuss the different justifications and implications that each of the two freedoms mentioned above have and how they enter such generalization. In order to accomplish this task, we begin by summarizing the setup of the problem in Section~\ref{Sec:CEsolution}, including the relativistic Boltzmann equation and the corresponding Chapman-Enskog first-order solution. In Section~\ref{Generalsolution} we write the general solution to the first-order integral equation, as formally derived in Ref.~\cite{Hydrolimit}, and describe each of the contributions to it in view of the choices of frame and representation, which are also briefly discussed. Section~\ref{frames} is devoted to the analysis of the homogeneous solution, where expressions for the coefficients involved are obtained in terms of general compatibility conditions. The particular cases of the so-called particle and energy frames are addressed in Section~\ref{Sect:particleenergy}. A summary and discussion of the results are included in Section~\ref{Discussion}, together with an analysis of the entropy production that clarifies the connection with non-equilibrium thermodynamics. We work in units for which the speed of light is one and consider a $(d+1)$-dimensional spacetime background whose metric $g_{\mu\nu}$ has signature $(-,+,\ldots,+)$.

\section{The Chapman-Enskog solution to the relativistic Boltzmann equation}
\label{Sec:CEsolution} 

In this section, we review the most important aspects of the relativistic Boltzmann equation for a dilute gas whose molecules undergo binary elastic collisions, together with the first-order solution corresponding to the CE expansion. See~\cite{wI63,Groot-Book,CercignaniKremer-Book,Vereshchagin-Book} for original references and textbooks on these subjects. The notation and procedure follows the formal description discussed in Ref.~\cite{Hydrolimit}. Here we only briefly quote the equations and results required for the discussion that follows, and the reader is referred to \cite{Hydrolimit} for further details.

\subsection{The relativistic Boltzmann equation}

The relativistic Boltzmann equation, 
\begin{equation}
L_{F}\left[f\right]=Q\left[f,f\right],
\label{BE}
\end{equation}
describes the evolution of the distribution function $f(x,p)$ 
by balancing
drift (left-hand side) and the effects of binary interactions (right-hand side) in phase space. Here, in terms of adapted local coordinates $\left(x^{\mu},p_{\mu}\right)$ on the cotangent bundle associated with the spacetime manifold,
\begin{equation}
L_{F}=p^{\mu}\frac{\partial}{\partial x^{\mu}}-\frac{1}{2}\frac{\partial g^{\alpha\beta}}{\partial x^{\mu}}p_{\alpha}p_{\beta}\frac{\partial}{\partial p_{\mu}}+qF_{\mu}{}^{\nu}p_{\nu}\frac{\partial}{\partial p_{\mu}},
\end{equation}
denotes the Liouville operator, including an external electromagnetic field  $F^{\mu\nu}$, and $Q[f,f](x,p_{1})$ is the integral operator that accounts for the balance of particles in and out of phase space cells due to binary elastic collisions. The particular
structure of $Q[f,f](x,p_{1})$  is not required for the purpose of the present work, however it is convenient to point out as one of its most important properties that 
\begin{equation}
\int\limits_{P_{x}^{+}(m)}\Psi\left(x,p\right)Q\left[f,f\right]\left(x,p\right)\dvol_{x}(p)=0,
\label{transfer}
\end{equation}
when $\Psi(x,p)\in\mathrm{span}\{1,p^{\mu}\}$ is a collision invariant. Here $\dvol_{x}(p)$ is
the Lorentz-invariant volume element and the integration is performed within the future mass hyperboloid $P_{x}^{+}(m)$~\cite{rAcGoS2022}.
Defining the particle current and energy-momentum-stress tensor as
\begin{equation}
J^{\mu}\left(f\right):=\int\limits_{P_{x}^{+}(m)}f\left(x,p\right)p^{\mu}\dvol_{x}(p),\qquad
T^{\mu\nu}\left(f\right)  :=\int\limits_{P_{x}^{+}(m)}f\left(x,p\right)p^{\mu}p^{\nu}\dvol_{x}(p),
\end{equation}
one obtains that Eqs.~(\ref{BE}) and (\ref{transfer}), together with the identities\footnote{See for instance Theorem~4 in~\cite{rAcGoS2022}.}
\begin{eqnarray}
\nabla^{\mu}J_{\mu}\left(f\right) & = & \int\limits_{P_{x}^{+}(m)}L_{F}\left[f\right]\dvol_{x}(p),\label{Eq:DivIdentity1}\\
\nabla^{\mu}T_{\mu\nu}\left(f\right)+qF^{\mu}{}_{\nu}J_{\mu}\left(f\right) & = & \int\limits_{P_{x}^{+}(m)}p_{\nu}L_{F}\left[f\right]\dvol_{x}(p),\label{Eq:DivIdentity2}
\end{eqnarray}
lead to the transport equations 
\begin{equation}
\nabla_{\mu}J^{\mu}=0,\qquad\nabla_{\mu}T^{\mu\nu}+qJ_{\mu}F^{\mu\nu}=0.\label{Eq:Hydro}
\end{equation}
Hence, particle and energy-momentum balances are direct consequences of the symmetries involved in the relativistic Boltzmann equation. Moreover, as widely known (see for example Refs.~\cite{rAcGoS2022,Groot-Book,CercignaniKremer-Book}) the local equilibrium configurations are given by the Jüttner distribution function: 
\begin{equation}
f^{(0)}\left(x,p\right)=\frac{n\left(x\right)}{2m\left(2\pi mk_{B}T\left(x\right)\right){}^{\frac{d-1}{2}}\bessel_{\frac{d+1}{2}}(z)}\exp\left[\frac{u^{\nu}\left(x\right)p_{\nu}}{k_{B}T\left(x\right)}\right],
\label{Eq:f0nTu}
\end{equation}
where $\bessel_{\ell}\left(z\right)$ denotes the modified Bessel function of the second kind of order $\ell$ and the argument $z:=m/k_{B}T$ measures the ratio between the rest energy of the individual particles and the thermal energy of the gas. The state variables that parametrize this distribution function are the particle number density $n$, the hydrodynamic velocity $u^{\mu}$ and the temperature $T$. In an equilibrium situation,
the particle current and energy-momentum-stress tensor have the form 
\begin{equation}
J^{\mu}=nu^{\mu},\qquad T^{\mu\nu}=neu^{\mu}u^{\nu}+\p\Delta^{\mu\nu},\label{JTeq}
\end{equation}
where $\Delta_{\nu}^{\mu}:=\delta_{\nu}^{\mu}+u^{\mu}u_{\nu}$ is the projector orthogonal to $u^\mu$, $e=e\left(T\right)$ is the internal energy density per unit mass, and $\p=\p(n,T)$ is the hydrostatic pressure. The balance equations~(\ref{Eq:Hydro}), for these equilibrium states, are given by the relativistic Euler equations which, denoting the space and time projections of $\nabla^{\mu}$ as $D^{\mu}=\Delta^{\mu\nu}\nabla_{\nu}$ and $\dot{\left(\right)}=u^{\mu}\nabla_{\mu}$ respectively, can be written as 
\begin{equation}
\dot{n}+\theta n=0,\quad\dot{T}+\frac{k_{B}}{c_{v}}\theta T=0,\quad a_{\nu}+\frac{D_{\nu}\p}{nh}-\frac{q}{h}E_{\nu}=0.
\end{equation}
Here, $\theta := \nabla^\mu u_\mu$, $a_\mu := \dot{u}_\mu$, $c_{v}:=\partial e/\partial T$, and $E_\nu := F_{\nu\mu} u^\mu$ is the electric field measured by a comoving observer.

\subsection{Chapman-Enskog first-order solution}

The CE expansion is motivated by the observation that a disparity between microscopic and macroscopic characteristic length scales leads to an ordering scheme for the Boltzmann equation such that it can be treated perturbatively. More precisely, if the characteristic macroscopic scale corresponding to the size of the system or to the distance in which the gradients of the state variables are significant is denoted by $\lms$ and the mean free path between collisions by $\lmfp$, for weak electromagnetic fields one finds that the right-hand side of Eq.~(\ref{BE}) differs from the left-hand side in one order of Knudsen's parameter $\varepsilon:=\lmfp/\lms$. In fact Eq.~(\ref{BE}) when written in terms of dimensionless quantities, yields 
\begin{equation}
L_{F}\left[f\right]=\frac{1}{\varepsilon}Q\left[f,f\right],
\end{equation}
which motivates the proposal of a perturbative solution as an expansion in terms of $\varepsilon$. Thus, one assumes that the non-equilibrium distribution function can be written in general as 
\begin{equation}
f=\sum_{\ell=0}^{\infty}\varepsilon^{\ell}f^{\left(\ell\right)},\label{CEexpanssion}
\end{equation}
where $f^{(0)}$ denotes a local Jüttner distribution as given in Eq.~(\ref{Eq:f0nTu}) which is parametrized by the variables $n$, $u^{\mu}$, and $T$ whose dynamics are determined by the transport equations to all orders in $\varepsilon$. That is, the first term in the expansion does not describe a global equilibrium configuration but rather the corresponding local equilibrium distribution function parametrized by out-of equilibrium state variables. In the present work, we are interested in analyzing the first-order correction within this method, and thus we seek a solution of the form (in terms of dimensional variables)
\begin{equation}
f\approx f^{(0)}+f^{(1)},\label{CE1storder}
\end{equation}
where it is understood that $f^{(1)}$ is of first order in $\varepsilon$ and $f^{(0)}$ is parametrized by $n$, $u^\nu$, and $T$ given by the equations of motion~(\ref{Eq:Hydro}) up to order 1 in $\varepsilon$. Introducing Eq.~(\ref{CE1storder}) into Eq.~(\ref{BE}) one obtains, to order zero 
\begin{equation}
Q\left[f^{(0)},f^{(0)}\right]=0,
\end{equation}
which is satisfied by the local Jüttner distribution function (\ref{Eq:f0nTu}). The next-order equation reads 
\begin{equation}
\mathcal{L}\left[f^{(1)}/f^{(0)}\right]=-L_{F}\left[\ln f^{(0)}\right],\label{1storder}
\end{equation}
where $\mathcal{L}$ is the linearized collision operator: 
\begin{equation}
\mathcal{L}\left[\phi\right]\left(x,p_{1}\right):=\int\limits_{P_{x}^{+}(m)}\int\limits_{S^{d-1}}\mathcal{F}\frac{d\sigma}{d\Omega}f_{2}^{(0)}\left[\phi_{1}+\phi_{2}-\phi_{1}^{*}-\phi_{2}^{*}\right]d\Omega\left(\hat{q}^{*}\right)\dvol_{x}(p_{2}).\label{LinearizedCO}
\end{equation}
Here $\mathcal{F}$ denotes the invariant flux, $d\sigma/d\Omega$ the differential cross section, and $\Omega$ the solid angle. The
subindex 1 or 2 is introduced in order to distinguish the colliding
particles, and the notation $(...)^{*}$ to differentiate between pre- and post-collision quantities.

\section{General first-order solution}

\label{Generalsolution} As thoroughly explained in Ref.~\cite{Hydrolimit}, in the presence of a weak background electromagnetic field,
the general first-order solution to Eq.~(\ref{1storder}) using the projection method can be written
as 
\begin{equation}
f^{(1)}=-f^{(0)}\left[\widehat{\mathcal{P}}_{\mu\nu}\mathcal{L}^{-1}\left[\left(p^{\mu}p^{\nu}\right)^{\perp}\right]+mA+B_{\mu}p^{\mu}\right],
\label{eq:f1}
\end{equation}
where: 
\begin{itemize}
\item $\left(...\right)^{\perp}$ denotes the orthogonal projection onto the orthogonal complement of $\ker\mathcal{L}=\mathrm{span}\{1,p^\mu\}$ with respect to the inner product
\begin{equation}
\langle\psi,\phi\rangle:=\int_{P_x^+(m)}\psi\left(p\right)\phi\left(p\right)f^{(0)}\left(x,p\right)\dvol_x(p),\label{innerproduct}
\end{equation}
see Section~V in Ref.~\cite{Hydrolimit} for details.
\item $\widehat{\mathcal{P}}_{\mu\nu}$ is given by
\begin{align}
\widehat{\mathcal{P}}_{\mu\nu} & :=\frac{1}{k_{B}T}\left\{ \sigma_{\mu\nu}-u_{\mu}\left(a_{\nu}+\frac{D_{\nu}T}{T}\right)+\left(\frac{\dot{T}}{T}+\frac{\theta}{d}\right)u_{\mu}u_{\nu}\right.\nonumber \\
 & \left.+\left[\hat{\Gamma}_{0}\left(\frac{\dot{n}}{n}+\theta\right)+\hat{\Gamma}_{1}\left(\frac{c_{v}}{k_{B}}\frac{\dot{T}}{T}+\theta\right)\right]u_{\mu}u_{\nu}+\hat{\Gamma}_{2}\left(\frac{h}{k_{B}T}a_{\nu}+\frac{D_{\nu}\p}{\p}-\frac{q}{k_{B}T}E_\nu\right)u_{\mu}\right\} ,\label{Eq:Phatmunu}
\end{align}
where $\sigma_{\mu\nu}$ and $h$ denote the shear and enthalpy per particle, respectively. The first line in Eq.~(\ref{Eq:Phatmunu}) corresponds to $\left(L_{F}\left[\ln f^{\left(0\right)}\right]\right)^{\perp}$. The second line adds a multiple of quantities that are of higher order in Knudsen's parameter. The freedom of including these terms arises from the fact that, due to the balance equations, the expressions in parentheses in the second line of Eq.~(\ref{Eq:Phatmunu}) can be added to the first-order solution without altering it to this order.
\item $\mathcal{L}^{-1}: (\ker\mathcal{L})^{\perp}\to (\ker\mathcal{L})^{\perp}$ denotes the inverse of the linearized collision operator when restricted to $(\ker\mathcal{L})^{\perp}$. Sufficient conditions on the cross section for the invertibility have been thoroughly discussed in Ref.~\cite{Hydrolimit}.
\item The last two terms in Eq.~(\ref{eq:f1}) correspond to an arbitrary function belonging $\ker\mathcal{L}$, which constitutes
the homogeneous solution. The quantities $A$ and $B_{\mu}$ are functions of the state variables only, and as will be explained in detail in the remaining of this article, they give rise to a family of constitutive relations.
\end{itemize}
It is important to point out that there is a profound difference in the arguments behind the introduction of the coefficients $A$ and $B_{\mu}$, and the ``parameters'' $\hat{\Gamma}_{0,1,2}$. The former are mandatory in order to obtain the most general solution to the integral equation and correspond to the homogeneous part of the solution. The latter parametrize the representation freedom and have nothing to do with the existence nor the uniqueness of the solution. At this point, it is worthwhile to emphasize that these terms are introduced as on-shell corrections within the microscopic framework, and they do not alter the frame choice. Furthermore, in the present approach, these terms appear in such a way that the TFP frame, which is obtained when $A$ and $B_{\mu}$ vanish and only the particular solution is present, is still exactly enforced to first order even off-shell. This is in contrast to phenomenological approaches where the frame is retained only on-shell up to first order (see for example~\cite{Bemfica2022}).

\section{Homogeneous solution and frame fixing}
\label{frames}

In order to write a general solution to the linearized Boltzmann equation
to first order within the Chapman-Enskog scheme, one first guarantees
existence of such solution by projecting the right-hand side of Eq.~(\ref{1storder}) onto the orthogonal complement of $\ker\mathcal{L}$.
The arguments that support the previous statement are thoroughly discussed
in Ref.~\cite{LaurePaper} for the non-relativistic case and in Ref.~\cite{Hydrolimit} for the relativistic generalization. A comparison of this method for assuring existence with the traditional one can be found in Ref.~\cite{JNET24}. Once the right-hand side is projected and the operator
can be inverted, the general solution must include an arbitrary element
of $\ker\mathcal{L}$, which leads to a family of solutions. In order to attain uniqueness, from the mathematical point of view, one needs $d+2$ conditions in order to fix the arbitrary coefficients $A$ and $B_{\mu}$. These conditions can be written in general form as 
\begin{equation}
\int\limits_{P_{x}^{+}(m)}\left(\begin{array}{c}
g_{1}\left(\gamma\right)\\
g_{2}\left(\gamma\right)\\
g_{3}\left(\gamma\right)\Delta_{\mu\nu}p^{\nu}
\end{array}\right)f^{(1)}\left(x,p\right)\dvol_x(p)=0,
\label{generalsubsidiary}
\end{equation}
where $\gamma:=-u_{\mu}p^{\mu}/m$. For convenience, from here on, we write $B_\mu = b_\mu - b u_\mu$ with $b_\mu :=\Delta_\mu^\alpha B_\alpha$ and $b := u^\mu B_\mu$, and require that Eq.~(\ref{generalsubsidiary}) uniquely determines the two scalar coefficients $A$ and $b$, and the vector quantity $b_{\mu}$. Notice that the equations in (\ref{generalsubsidiary}) reduce to the compatibility constraints introduced in Ref.~\cite{Rocha2022} when $g_{1,2}(\gamma)=\gamma^{q,s}$ and $g_{3}(\gamma)=\gamma^{z}$ (where $q$, $s$, and $z$ correspond to the notation used in~\cite{Rocha2022}). However, in the present work, $g_{i}$ can in principle be any function of $\gamma$ as long as $g_1$ and $g_2$ are independent, the integrals
in Eq.~(\ref{generalsubsidiary}) exist and the denominators in Eqs.~(\ref{Ag}-\ref{Cg}) below do not vanish.

Introducing the solution given in Eq.~(\ref{eq:f1}) into Eq.~(\ref{generalsubsidiary}) one finds that the coefficients can be written as
\begin{equation}
mA=\widehat{\mathcal{P}}_{\mu\nu}\mathcal{A}^{\mu\nu},\qquad mb=\widehat{\mathcal{P}}_{\mu\nu}\mathcal{B}^{\mu\nu},\qquad b_{\alpha}=\widehat{\mathcal{P}}_{\mu\nu}\mathcal{C}_{\alpha}\,^{\mu\nu},
\label{eq:Abb}
\end{equation}
where \footnote{Note that $\left\langle g_i,p^\mu \right\rangle$ is proportional to $u^\mu$ and $\Delta^\mu{}_\alpha\Delta^\nu{}_\beta\left\langle g_i,p^\alpha p^\beta \right\rangle$ to $\Delta^{\mu\nu}$.}
\begin{align}
\mathcal{A}^{\mu\nu} & :=-\frac{\left\langle g_{1},\mathcal{L}^{-1}\left[\left(p^{\mu}p^{\nu}\right)^{\perp}\right]\right\rangle \left\langle g_{2},\gamma\right\rangle -\left\langle g_{2},\mathcal{L}^{-1}\left[\left(p^{\mu}p^{\nu}\right)^{\perp}\right]\right\rangle \left\langle g_{1},\gamma\right\rangle }{\left\langle g_{1},1\right\rangle \left\langle g_{2},\gamma\right\rangle -\left\langle g_{1},\gamma\right\rangle \left\langle g_{2},1\right\rangle },\label{Ag}\\
\mathcal{B}^{\mu\nu} & :=-\frac{\left\langle g_{2},\mathcal{L}^{-1}\left[\left(p^{\mu}p^{\nu}\right)^{\perp}\right]\right\rangle \left\langle g_{1},1\right\rangle -\left\langle g_{1},\mathcal{L}^{-1}\left[\left(p^{\mu}p^{\nu}\right)^{\perp}\right]\right\rangle \left\langle g_{2},1\right\rangle }{\left\langle g_{1},1\right\rangle \left\langle g_{2},\gamma\right\rangle -\left\langle g_{1},\gamma\right\rangle \left\langle g_{2},1\right\rangle},\label{Bg}\\
\mathcal{C}_{\alpha}\,^{\mu\nu} & :=-\frac{d}{m^{2}}\frac{\Delta_{\alpha\beta}\left\langle g_{3}p^{\beta},\mathcal{L}^{-1}\left[\left(p^{\mu}p^{\nu}\right)^{\perp}\right]\right\rangle }{\left\langle g_{3}\gamma,\gamma\right\rangle -\left\langle g_{3},1\right\rangle }.\label{Cg}
\end{align}
Also note that we have suppressed the $\gamma$-dependence of $g_{i}$
within the inner product, and we will do so throughout the rest of this work in order to simplify the notation.

The freedom of choosing $g_{1}$, $g_{2}$, and $g_{3}$ has important implications. The fact that any linear combination of collision invariants in the solution~(\ref{eq:f1}) is valid is directly related to the freedom of choosing a frame, as already discussed above.  That is, in equilibrium the variables involved in $J^{\mu}$ and $T^{\mu\nu}$ are unambiguously associated with the state variables $n$, $u^{\mu}$, and $T$ (see Eq.~(\ref{JTeq})). However, there is no unique way to describe non-equilibrium states and in principle any choice is valid as long as the requirement that all of them coincide in equilibrium with Eq.~(\ref{JTeq}) is fulfilled. The way in which that happens characterizes the particular  frame and thus, since Eq.~(\ref{generalsubsidiary}) implies 
\begin{align}
\int\limits_{P_{x}^{+}(m)}g_{1}\left(\gamma\right)f\left(x,p\right)\dvol_x(p) & =\int\limits_{P_{x}^{+}(m)}g_{1}\left(\gamma\right)f^{\left(0\right)}\left(x,p\right)\dvol_x(p),
\label{eq:s1}\\
\int\limits_{P_{x}^{+}(m)}g_{2}\left(\gamma\right)f\left(x,p\right)\dvol_x(p) & =\int\limits_{P_{x}^{+}(m)}g_{2}\left(\gamma\right)f^{\left(0\right)}\left(x,p\right)\dvol_x(p),
\label{eq:s2}\\
\int\limits_{P_{x}^{+}(m)}g_{3}\left(\gamma\right)\Delta_{\mu\nu}p^{\nu}f\left(x,p\right)\dvol_x(p) & =\int\limits_{P_{x}^{+}(m)}g_{3}\left(\gamma\right)\Delta_{\mu\nu}p^{\nu}f^{\left(0\right)}\left(x,p\right)\dvol_x(p),
\label{eq:s3}
\end{align}
the quantities that are anchored to their equilibrium values are the ones on the left-hand sides of Eqs.~(\ref{eq:s1})-(\ref{eq:s3}) and determine the particular frame. In this sense, the functions $g_{i}$ determine conditions which are the equivalent of the matching conditions involved in the general first-order theories for dissipative fluids.

In order to explicitly write $f^{(1)}$ in terms of the driving forces, we begin by defining 
\begin{equation}
\mathcal{R}_{\alpha}\,^{\mu\nu}\left[\varUpsilon\right]=\left\langle \varUpsilon\left(\gamma\right)p_{\alpha},\mathcal{L}^{-1}\left[\left(p^{\mu}p^{\nu}\right){}^{\perp}\right]\right\rangle ,\label{Eq:RDef-1}
\end{equation}
where $\varUpsilon$ is an arbitrary function of $\gamma$, such that the scalar product exists. Notice that, since $\mathcal{R}_{\alpha}\,^{\mu\nu}\left[\varUpsilon\right]$ is a Lorentz tensor which can only depend on the equilibrium quantities and satisfies $g_{\mu\nu}\mathcal{R}_{\alpha}\,^{\mu\nu}\left[\varUpsilon\right]=0$,
it is intuitively clear that it can be written as
\begin{equation}
\mathcal{R}_{\alpha}\,^{\mu\nu}\left[\varUpsilon\right]=R_{1}\left[\varUpsilon\right]\left(u^{\mu}u^{\nu}+\frac{1}{d}\Delta^{\mu\nu}\right)u_{\alpha}+R_{2}\left[\varUpsilon\right]\Delta_{\alpha}{}^{(\mu}u^{\nu)},
\label{Eq:RReduced-1}
\end{equation}
where the parentheses $(\ldots)$ denote symmetrization, $v_{(\mu\nu)} := (v_{\mu\nu} + v_{\nu\mu})/2$, and
\begin{equation}
R_{1}\left[\varUpsilon\right]=m^{3}\left\langle \varUpsilon\left(\gamma\right)\gamma,\mathcal{L}^{-1}\left[\left(\gamma^{2}\right){}^{\perp}\right]\right\rangle ,\qquad R_{2}\left[\varUpsilon\right]=\frac{2m}{d}\Delta_{\alpha\beta}\left\langle \varUpsilon\left(\gamma\right)p^{\beta},\mathcal{L}^{-1}\left[\left(\gamma p^{\alpha}\right){}^{\perp}\right]\right\rangle .
\label{Eq:R1R2Def}
\end{equation}
A proof of this fact which is based on the symmetries of the linearized collision is included in Appendix~\ref{App:RCoeff} for completeness. Using Eqs.~(\ref{Eq:RDef-1}) and (\ref{Eq:RReduced-1}), one obtains
\begin{align}
\mathcal{A}^{\mu\nu} =-\frac{\chi_1}{m}\left(u^{\mu}u^{\nu}+\frac{1}{d}\Delta^{\mu\nu}\right),\qquad
\mathcal{B}^{\mu\nu} =-\frac{\chi_2}{m}\left(u^{\mu}u^{\nu}+\frac{1}{d}\Delta^{\mu\nu}\right),\qquad 
\mathcal{C}_{\alpha}\,^{\mu\nu} =-\frac{d\chi_3}{m^{2}}\Delta_{\alpha}{}^{(\mu}u^{\nu)},
\end{align}
with $\chi_i$ defined as follows:
\begin{align}
\chi_{1} & :=\frac{R_{1}\left[g_{1}/\gamma\right]\left\langle g_{2},\gamma\right\rangle -R_{1}\left[g_{2}/\gamma\right]\left\langle g_{1},\gamma\right\rangle }{\left\langle g_{1},1\right\rangle \left\langle g_{2},\gamma\right\rangle -\left\langle g_{1},\gamma\right\rangle \left\langle g_{2},1\right\rangle },
\label{chi1}\\
\chi_{2} & :=\frac{R_{1}\left[g_{2}/\gamma\right]\left\langle g_{1},1\right\rangle -R_{1}\left[g_{1}/\gamma\right]\left\langle g_{2},1\right\rangle }{\left\langle g_{1},1\right\rangle \left\langle g_{2},\gamma\right\rangle -\left\langle g_{1},\gamma\right\rangle \left\langle g_{2},1\right\rangle },
\label{chi2}\\
\chi_{3} & :=\frac{R_{2}\left[g_{3}\right]}{\left\langle g_{3}\gamma,\gamma\right\rangle -\left\langle g_{3},1\right\rangle }\label{chi3}.
\end{align}
Since 
\begin{align}
\widehat{\mathcal{P}}_{\mu\nu}\left(u^{\mu}u^{\nu}+\frac{1}{d}\Delta^{\mu\nu}\right) & =\frac{1}{k_{B}T}\left[\hat{\Gamma}_{0}\frac{\dot{n}}{n}+\left(\frac{c_{v}}{k_{B}}\hat{\Gamma}_{1}+1\right)\frac{\dot{T}}{T}+\left(\hat{\Gamma}_{0}+\hat{\Gamma}_{1}+\frac{1}{d}\right)\theta\right],\\[2pt]
\widehat{\mathcal{P}}_{\mu\nu}\Delta_{\alpha}{}^{(\mu}u^{\nu)} & =\frac{1}{2k_{B}T}\left[\left(1-\frac{h}{k_{B}T}\hat{\Gamma}_{2}\right)a_{\alpha}+\left(1-\hat{\Gamma}_{2}\right)\frac{D_{\alpha}T}{T}-\hat{\Gamma}_{2}\frac{D_{\alpha}n}{n}+\hat{\Gamma}_{2}\frac{qE_{\alpha}}{k_{B}T}\right],
\end{align}
the coefficients involved in the homogeneous contribution in Eq.~(\ref{eq:f1}) can be written as 
\begin{equation}
A=-\frac{\Phi_{S}}{m^{2}k_{B}T}\chi_{1},\quad b=-\frac{\Phi_{S}}{m^{2}k_{B}T}\chi_{2},\quad b^{\alpha}=-\frac{d}{2}\frac{\Phi_{V}^{\alpha}}{m^{2}k_{B}T}\chi_{3},\label{Abbalpha}
\end{equation}
where we have defined
\begin{align}
\Phi_{S} & :=\hat{\Gamma}_{0}\frac{\dot{n}}{n}+\left(\frac{c_{v}}{k_{B}}\hat{\Gamma}_{1}+1\right)\frac{\dot{T}}{T}+\left(\hat{\Gamma}_{0}+\hat{\Gamma}_{1}+\frac{1}{d}\right)\theta,\\
\Phi_{V}^{\alpha} & :=\left(1-\frac{h}{k_{B}T}\hat{\Gamma}_{2}\right)a^{\alpha}+\left(1-\hat{\Gamma}_{2}\right)\frac{D^{\alpha}T}{T}-\hat{\Gamma}_{2}\frac{D^{\alpha}n}{n}+\hat{\Gamma}_{2}\frac{qE^{\alpha}}{k_{B}T}.
\end{align}
Note that, in Eqs.~(\ref{Abbalpha}) the information related with the frame is introduced through the $\chi_i$ coefficients while the representation freedom is contained in $\Phi_S^{\alpha} $ and $\Phi_V^{\alpha}$.

\section{Non-Equilibrium contributions}

In order to express the non-equilibrium corrections in a general fashion, we start by separating $J^{\mu}$ and $T^{\mu\nu}$ in their zeroth and first-order contributions:
\begin{align}
J^{\mu} & =J_{\left(0\right)}^{\mu}+J_{\left(1\right)}^{\mu},\qquad T^{\mu\nu}=T_{\left(0\right)}^{\mu\nu}+T_{\left(1\right)}^{\mu\nu},
\end{align}
where the non-equilibrium corrections are 
\begin{equation}
J_{\left(1\right)}^\mu := J^\mu\left(f^{(1)}\right),\quad T_{\left(1\right)}^{\alpha\beta} := T^{\alpha\beta}\left(f^{(1)}\right).
\end{equation}

Regarding $J_{\left(1\right)}^{\alpha}$, notice that since $\left\langle p^{\alpha},\mathcal{L}^{-1}\left[\left(p^{\mu}p^{\nu}\right){}^{\perp}\right]\right\rangle =0$,
the non-equilibrium contribution to the current only depends on the homogeneous solution. Indeed, using Eq. (\ref{eq:f1}) one obtains
\begin{equation}
J_{\left(1\right)}^{\alpha}=-\left[mA\left\langle 1,p^{\alpha}\right\rangle +mb\left\langle \gamma,p^{\alpha}\right\rangle +b_{\beta}\left\langle p^{\alpha},p^{\beta}\right\rangle \right].
\end{equation}
Defining the non-equilibrium contributions to $J^{\alpha}$
as 
\begin{equation}
n_{(1)}:=-u_{\alpha}J_{\left(1\right)}^{\alpha},\qquad\mathcal{J}^{\alpha}:=\Delta_{\beta}^{\alpha}J_{\left(1\right)}^{\beta},
\end{equation}
leads to
\begin{align}
n^{(1)} & =\frac{n}{mk_{B}T}\left(\chi_{1}+\frac{e}{m}\chi_{2}\right)\Phi_{S},
\label{eq:n1-1}\\
\mathcal{J}^{\alpha} & =\frac{dn}{2m^{2}}\chi_{3}\Phi_{V}^{\alpha}.
\label{eq:j-1}
\end{align}

Similarly, the first-order contribution to the energy-momentum-stress tensor can be written as
\begin{equation}
T_{\left(1\right)}^{\alpha\beta}=-\left[ \widehat{\mathcal{P}}_{\mu\nu}\mathcal{S}^{\alpha\beta\mu\nu}+mA\left\langle p^{\alpha}p^{\beta},1\right\rangle +mb\left\langle p^{\alpha}p^{\beta},\gamma\right\rangle +b_{\delta}\left\langle p^{\alpha}p^{\beta},p^{\delta}\right\rangle \right] ,
\end{equation}
where we have introduced the tensor $\mathcal{S}^{\alpha\beta\mu\nu}:=\langle p^{\alpha}p^{\beta},\mathcal{L}^{-1}\left[\left(p^{\mu}p^{\nu}\right){}^{\perp}\right]\rangle=\langle\left(p^{\alpha}p^{\beta}\right)^{\perp},\mathcal{L}^{-1}\left[\left(p^{\mu}p^{\nu}\right){}^{\perp}\right]\rangle$,
following Ref.~\cite{Hydrolimit}, which can be decomposed as 
\begin{align}
\mathcal{S}^{\alpha\beta\mu\nu} & =\mathcal{S}_{1}u^{\alpha}u^{\beta}u^{\mu}u^{\nu}+\mathcal{S}_{2}\Delta^{\alpha\beta}\Delta^{\mu\nu}+\frac{1}{2}\mathcal{S}_{3}\left(\Delta^{\alpha\mu}\Delta^{\beta\nu}+\Delta^{\beta\mu}\Delta^{\alpha\nu}\right)\nonumber \\
 & +\frac{1}{2}\mathcal{S}_{4}\left(\Delta^{\alpha\beta}u^{\mu}u^{\nu}+\Delta^{\mu\nu}u^{\alpha}u^{\beta}\right)+\frac{1}{4}\mathcal{S}_{5}\left(\Delta^{\alpha\mu}u^{\beta}u^{\nu}+\Delta^{\alpha\nu}u^{\beta}u^{\mu}+\Delta^{\beta\mu}u^{\alpha}u^{\nu}+\Delta^{\beta\nu}u^{\alpha}u^{\mu}\right),\label{S}
\end{align}
due to its symmetry properties.\footnote{Notice that the definition of the scalar product $\langle.\rangle$ implies $\mathcal{S}^{\alpha\beta\mu\nu}=\mathcal{S}^{\mu\nu\alpha\beta}$. Also, one clearly has $\mathcal{S}^{\mu\nu\alpha\beta}=\mathcal{S}^{\nu\mu\alpha\beta}=\mathcal{S}^{\nu\mu\beta\alpha}$,
and since $1\in\ker\left(\mathcal{L}\right)$, $g_{\alpha\beta}\mathcal{S}^{\alpha\beta\mu\nu}=g_{\mu\nu}\mathcal{S}^{\alpha\beta\mu\nu}=0$,
such that $\frac{1}{d}\mathcal{S}_{1}=\frac{1}{2}\mathcal{S}_{4}=d\mathcal{S}_{2}+\mathcal{S}_{3}$. Hence we can take $\mathcal{S}_1$, $\mathcal{S}_3$, and $\mathcal{S}_5$ as independent coefficients.} If we define the non-equilibrium contributions to $T_{\left(1\right)}^{\alpha\beta}$
as  
\begin{eqnarray}
\begin{split}
ne^{\left(1\right)} & :=u_{\alpha}u_{\beta}T_{\left(1\right)}^{\alpha\beta},\qquad & \p^{\left(1\right)} & :=\frac{1}{d}\Delta_{\alpha\beta}T_{\left(1\right)}^{\alpha\beta},\\
\mathcal{Q}^{\lambda} & :=-u_{\alpha}\Delta_{\beta}^{\lambda}T_{\left(1\right)}^{\alpha\beta},\qquad & \mathcal{T^{\lambda\sigma}} & :=\left( \Delta_{\alpha}^{\lambda}\Delta_{\beta}^{\sigma} -\frac{1}{d}\Delta_{\alpha\beta}\Delta^{\lambda\sigma}\right)T_{\left(1\right)}^{\alpha\beta},
\end{split}
\end{eqnarray}
one obtains \footnote{The following expressions are useful: $\left\langle 1,1 \right \rangle = \frac{n}{m} G_{-1}$,  $\left\langle 1,\gamma \right \rangle = \frac{n}{m}$,  $\left\langle 1,\gamma^2 \right \rangle = \frac{n e}{m^2}$,  $\left\langle 1,\gamma^3 \right \rangle = \frac{n}{m} (G_2 - 3G_1/z)$.}
\begin{align}
ne^{\left(1\right)} & =-\frac{1}{k_{B}T}\left[\mathcal{S}_{1}-\frac{ne}{m}\chi_{1}+n\left(\frac{3}{z}G_{1}-G_{2}\right)\chi_{2}\right]\Phi_{S},\label{eq:e1-1}\\
\p^{\left(1\right)} & =-\frac{1}{k_{B}T}\left[\frac{1}{d}\mathcal{S}_{1}-\frac{\p}{m}\left(\chi_{1}+\chi_{2}G_{1}\right)\right]\Phi_{S},\label{eq:p1-1}\\
\mathcal{Q}^{\alpha} & =-\frac{1}{k_{B}T}\left[\frac{1}{4}\mathcal{S}_{5}-\frac{d}{2m}\chi_{3}\p G_{1}\right]\Phi_{V}^{\alpha},\label{eq:q-1}\\
\mathcal{T^{\alpha\beta}}  & =-\frac{\mathcal{S}_{3}}{k_B T}\sigma^{\alpha\beta},\label{eq:tau-1}
\end{align}
where we have introduced the auxiliary functions $G_{k}$ defined as $G_{k}(z):=\textbf{K}_{\frac{d+1}{2}+k}(z)/\textbf{K}_{\frac{d+1}{2}}(z)$, $k=-1,0,1,2,\ldots$ whose properties and recursion relations can be consulted, for example, in Refs.~\cite{CercignaniKremer-Book,Hydrolimit}. Thus, one can write in general, for the particle current and energy-momentum-stress tensor describing the relativistic dissipative fluid, 
\begin{align}
J^{\mu} & =\left(n+n^{(1)}\right)u^{\mu}+\mathcal{J^{\mu}},\\
T^{\mu\nu} & =n\left(e+e^{\left(1\right)}\right)u^{\mu}u^{\nu}+\left(\p+\p^{\left(1\right)}\right)\Delta^{\mu\nu}+2u^{(\mu}\mathcal{Q}^{\nu)}+\mathcal{T^{\mu\nu}},
\end{align}
with the constitutive equations given by Eqs.~(\ref{eq:n1-1}), (\ref{eq:j-1})
and (\ref{eq:e1-1})-(\ref{eq:tau-1}). Furthermore, it is important to point out at this stage that, as shown in Ref.~\cite{Hydrolimit}, the coefficients $\mathcal{S}_{1,3,5}$ that appear in these equations are directly related to the three invariant transport coefficients namely, the thermal conductivity $\kappa$ and the bulk and shear viscosities  $\zeta$ and $\eta$, respectively. More precisely, one has
\begin{equation}
\mathcal{S}_{1}=k_{B}T\left(\frac{1}{d}-\frac{k_{B}}{c_{v}}\right)^{-2}\zeta,\qquad\mathcal{S}_{5}=4k_{B}T\kappa,\qquad\mathcal{S}_{3}=2k_{B}T\eta.
\label{eses}
\end{equation}

\section{Solution in particular frames}
\label{Sect:particleenergy}

In the present work, as well as in the most
recent literature on first-order relativistic fluid theories, we adopt
the term \textit{frame} as the
set of variables that describe the non-equilibrium state by specifying
the way in which they converge to the state variables $n$, $T$,
and $u^{\mu}$ in equilibrium. This choice is determined by fixing
matching conditions in the phenomenological approach and compatibility
conditions (choosing $g_{i}$ in Eq. (\ref{generalsubsidiary})) within
kinetic theory. However, this terminology is unfortunate, as was already pointed out in Ref.~\cite{pKovtun19}, and can be misleading since one usually associates the term frame to a \textit{reference} frame. Such term may have been adopted since the first and most traditional descriptions for dissipative fluids, proposed by Eckart~\cite{cE1940} and Landau~\cite{LandauLifshitz-Book6}, only differ from each other in the
choice for the hydrodynamic velocity and thus, can be misinterpreted
as considering two different observers for the description of the
system. Indeed, for both cases one has, using the non-equilibrium
quantities defined in the previous section: 
\begin{equation}
e_{E}^{\left(1\right)}=e_{L}^{\left(1\right)}=0,\quad n_{E}^{\left(1\right)}=n_{L}^{\left(1\right)}=0.
\end{equation}
The velocity in the case of Eckart is fixed by imposing $J_{\left(1\right)E}^{\mu}=0$
while the Landau proposal is characterized by $T_{\left(1\right)L}^{\mu\nu}u_{\mu L}=0$.
It is probably because of these particular choices that the two versions
of linear relativistic non-equilibrium thermodynamics were referred
to as two different frames. However, it is important to remark that in Refs.~\cite{wI76,wIjS79a} W. Israel and J. M. Stewart recognized the possibility of considering other frames as first-order corrections from equilibrium configurations by performing first-order transformations on the velocity. This gave rise to a family of allowed frames, which only differ in the velocity and for which the theory remains invariant, reinforcing the idea that a frame is directly linked to an observer. Nevertheless, in the present context the term frame is broader and refers to the assignment of all thermodynamic variables and not only the hydrodynamic velocity.

Having clarified this point, in the rest of this section we obtain the constitutive equations for particular choices of two of the functions $g_{i}$ leaving the third one unspecified, which leads to a family of frames. In this sense, we define a particle frame (following Refs.~\cite{aGjSoS2024a},\cite{aGjSoS2024b},\cite{JNET24}) as any frame for which $J^{\mu}=nu^{\mu}$ holds even out of equilibrium, and an energy frame as any frame for which $T_{\left(1\right)}^{\mu\nu}u_{\mu}=0$. The particular cases of the TFP and Eckart's (particle) and Landau's (energy) frames are addressed for illustrative purposes.

\subsection{Coefficients in a particle frame}

As mentioned above, we first consider frames for which $J^\mu_{(1)} = 0$. This condition corresponds to choosing $g_{2}=\gamma$ and $g_{3}=1$, leaving $g_1$ unspecified. In this case, $R_1[g_2/\gamma] = R_2[g_3] = 0$ and Eqs.~(\ref{chi1})-(\ref{chi3}) yield
\begin{equation}
\chi_{1}=-\frac{e}{m}\chi_{2}=\frac{R_{1}\left[g_{1}/\gamma\right]\frac{e}{m}}{\left\langle g_{1},1\right\rangle \frac{e}{m}-\left\langle g_{1},\gamma\right\rangle }\quad \textrm{and}\quad \chi_3=0.\label{ABPF}
\end{equation}
Introducing these results into Eqs.~(\ref{eq:n1-1}) and (\ref{eq:j-1}) leads to $n^{(1)}=0$ and $\mathcal{J}^{\alpha}=0$, which is consistent with the condition $J^{\mu}=nu^{\mu}$
in particle frames. For the remaining non-equilibrium quantities, Eqs.~(\ref{eq:e1-1})-(\ref{eq:tau-1}) lead to
\footnote{It is useful to consider the relations $\left(m^{2}/e^{2}\right)\left(\frac{3}{z}G_{1}-G_{2}\right)+1=-\left(m^{2}/e^{2}\right)c_{v}/\left(z^{2}k_{B}\right)$,
$e/m=G_{1}-z^{-1}$ and $c_{v}/k_{B}=z^{2}+(d+2)zG_{1}-z^{2}G_{1}^{2}-1$.}
\begin{align}
 ne^{\left(1\right)}&=-\frac{1}{k_{B}T}\left(\mathcal{S}_{1}+\frac{nm}{z^{2}e}\frac{c_{v}}{k_{B}}\chi_{1}\right)\Phi_{S},\\ \p^{\left(1\right)}&=-\frac{1}{k_{B}T}\left(\frac{1}{d}\mathcal{S}_{1}+\frac{\p}{ze}\chi_{1}\right)\Phi_{S},\\
 \mathcal{Q}^{\alpha}&=-\frac{\mathcal{S}_{5}}{4k_{B}T}\Phi_{V}^{\alpha}.\label{eq:particleQ} 
\end{align}

In the particular case of the TFP frame $g_1=1$, implying $\chi_1=0$ and $ne^{(1)}=d\mathrm{p}^{(1)}$. In contrast, in Eckart's frame~\cite{cE1940} for which $g_{1}=\gamma^{2}$, one has $R_1[g_1/\gamma] = \mathcal{S}_1/m$, which leads to 
\begin{equation}
\chi_{1}=-\mathcal{S}_{1}\frac{z^{2}k_{B}}{nc_{v}}\frac{e}{m},
\end{equation}
such that $e^{\left(1\right)}=0$, and
\begin{equation}
\p^{\left(1\right)}=-\frac{\mathcal{S}_{1}}{k_{B}T}\left(\frac{1}{d}-\frac{k_{B}}{c_{v}}\right)\Phi_{S}.
\label{eq:p1E}
\end{equation}
In Eckart's frame one then recovers the well-known result 
\begin{equation}
J^{\mu}  =nu^{\mu},\quad
T^{\mu\nu}  =neu^{\mu}u^{\nu}+\left(\p+\p^{\left(1\right)}\right)\Delta^{\mu\nu}+2u^{(\mu}\mathcal{Q}^{\nu)}+\mathcal{T^{\mu\nu}},
\end{equation}
where $\p^{\left(1\right)}$, $\mathcal{Q}^{\mu}$, and $\mathcal{T^{\mu\nu}}$
are given by Eqs.~(\ref{eq:p1E}), (\ref{eq:particleQ}) and (\ref{eq:tau-1}).
In order to obtain the usual constitutive relations one considers $\hat{\Gamma}_{0}=\hat{\Gamma}_{2}=0$ and $\hat{\Gamma}_{1}=-k_{B}/c_{v}$ which eliminates the time derivatives of the state variables in $\Phi_S$. This is carefully explained in Section~IV of Ref.~\cite{JNET24} and corresponds to a change of representation within Eckart's frame. Moreover, introducing the relations given in Eq.~(\ref{eses}) one obtains the well-known expressions:
\begin{align}
\p^{\left(1\right)}=-\zeta\theta \quad \text{and}\quad
\mathcal{Q}^{\alpha}=-\kappa\left(\frac{D_{\alpha}T}{T}+a_{\alpha}\right).
\end{align}

\subsection{Coefficients in an energy frame}

In order to explore the structure of the constitutive equations under the condition $u_{\mu}T_{(1)}^{\mu\nu} = 0$, we need to choose $g_2=\gamma^2$ and $g_3=\gamma$, leaving  $g_{1}$ unspecified. We adopt the nomenclature for this family of frames as energy frames,
based on Landau's theory~\cite{LandauLifshitz-Book6}. However, notice that Landau's frame also fixes $g_{1}=\gamma$, which will only be incorporated at the end of this subsection. 

These choices of $g_{2,3}$ lead to $R_{1}\left[g_{2}/\gamma\right]=\mathcal{S}_{1}/m$ and $R_{2}\left[g_{3}\right]=\mathcal{S}_{5}/\left(2m\right)$, and thus 
\begin{align}
\chi_{1}=\frac{m}{ne}\left[ \mathcal{S}_{1}+n\left(\frac{3}{z}G_{1}-G_{2}\right)\chi_{2}\right]\quad\text{and}\quad\chi_{3}=\frac{m}{2\text{p}}\frac{\mathcal{S}_{5}}{dG_{1}},
\end{align}
from which one obtains $ne^{(1)}=0$ and $\mathcal{Q}^{\mu}=0$ (see Eqs.~(\ref{eq:e1-1}) and (\ref{eq:q-1})). The rest of the constitutive equations, other than $\mathcal{T}^{\mu\nu}$, are still subject to the choice of $g_{1}$, and they can be written as 
\begin{align}
n^{(1)} & =\frac{1}{e k_{B}T}\left(\mathcal{S}_{1}-\frac{n}{z^{2}}\frac{c_{v}}{k_{B}}\chi_{2}\right)\Phi_{S},\\
\p^{\left(1\right)} & =-\frac{1}{k_{B}T}\left\{ \frac{\mathcal{S}_{1}}{d}-\frac{\p}{e}\left[\frac{\mathcal{S}_{1}}{n}-\frac{\chi_{2}}{z^{2}}\left(\frac{c_{v}}{k_{B}}-\frac{ze}{m}\right)\right]\right\} \Phi_{S},\\
\mathcal{J}^{\alpha} & =\frac{1}{4mk_{B}T}\frac{\mathcal{S}_{5}}{G_{1}}\Phi_{V}^{\alpha}.
\end{align}
Notice that $\mathcal{J}^{\mu}$ is proportional to the heat flux $\mathcal{Q}^\mu$ obtained in a particle frame (see Eq.~(\ref{eq:particleQ})). More precisely, one recovers the known result $\mathcal{J}^{\mu} = -\mathcal{Q}^{\mu}/h$ where $h=mG_{1}$ is the enthalpy per particle~\cite{CercignaniKremer-Book}. 

For the Landau frame \cite{LandauLifshitz-Book6}, one additionally fixes $g_{1}=\gamma$, such that $R_{1}\left[g_{1}/\gamma\right]=0$ and $\chi_2 = z^2\frac{k_B}{n c_v}\mathcal{S}_1$, and thus one recovers the results for the scalar quantities found in Eckart's frame namely, $n^{(1)}=0$ and $\p^{\left(1\right)}$ as given in Eq. (\ref{eq:p1E}). Notice that Eckart and Landau's frame only differ in the vector compatibility condition (\ref{eq:s3}) involving $g_{3}$, which fixes the hydrodynamic velocity $u^{\nu}$. As mentioned above, this fact may have led to the confusion of identifying changes of frames with changes of reference frames.

\section{Summary and discussion}
\label{Discussion}

In this work we have formally shown that the first-order out of equilibrium solution that is obtained using the projection method developed in Ref.~\cite{LaurePaper} can be generalized in order to establish constitutive equations that couple dissipative fluxes to all possible forces of the corresponding tensorial rank. These relations contain three coefficients that are related to the choice of frame $\chi_{i}$ and three arbitrary parameters $\hat{\Gamma}_{i}$ which correspond to the freedom of choosing a representation, and they can be written as 
\begin{align}
n^{\left(1\right)}=\sum_{i=1}^{3}\nu_{i}F^{i},\qquad ne^{\left(1\right)} & =\sum_{i=1}^{3}\varepsilon_{i}F^{i},\qquad\p^{\left(1\right)}=\sum_{i=1}^{3}\pi_{i}F^{i},\label{Eq:GeneralCR1}\\
\mathcal{J}^{\alpha}=\sum_{i=1}^{4}\gamma_{i}\bar{F}^{i\alpha},\qquad\mathcal{Q}^{\alpha} & =\sum_{i=1}^{4}\kappa_{i}\bar{F}^{i\alpha},\qquad\mathcal{T^{\alpha\beta}}=-2\eta\text{F}^{\alpha\beta},\label{Eq:GeneralCR2}
\end{align}
with $F^{1}=\dot{n}/n$, $F^{2}=\dot{T}/T$, $F^{3}=\theta$, $\bar{F}^{1\alpha}=D^{\alpha}n/n$,
$\bar{F}^{2\alpha}=D^{\alpha}T/T$, $\bar{F}^{3\alpha}=a^{\alpha}$,
$\bar{F}^{4\alpha}=qE^{\alpha}/h$, and $\text{F}^{\alpha\beta}=\sigma^{\alpha\beta}$.
The main result of this article is to have revealed the most general
structure of the coefficients $\nu_{i}$, $\varepsilon_{i}$, $\pi_{i}$,
$\gamma_{i}$, and $\kappa_{i}$ within the microscopic derivation
of the fluid theory. Namely, they have the particular product form
$\nu_{i}=c_{\nu}v_{i}$, $\varepsilon_{i}=c_{\varepsilon}v_{i}$,
$\pi_{i}=c_{\pi}v_{i}$, $\gamma_{i}=c_{\gamma}w_{i}$, $\kappa_{i}=c_{\kappa}w_{i}$,
where 
\begin{align}
c_{\nu} & =\frac{n}{mk_{B}T}\left(\chi_{1}+\frac{e}{m}\chi_{2}\right), & c_{\gamma} & =-\frac{dn}{2m^{2}}\chi_{3},\nonumber \\
c_{\varepsilon} & =-\frac{1}{k_{B}T}\left[\mathcal{S}_{1}-\frac{ne}{m}\left(\chi_{1}+\frac{e}{m}\chi_{2}\right)-\frac{n}{z^{2}}\frac{c_{v}}{k_{B}}\chi_{2}\right], & c_{\kappa} & =\frac{1}{k_{B}T}\left(\frac{1}{4}\mathcal{S}_{5}-\frac{d}{2}\frac{h\p}{m^{2}}\chi_{3}\right),\label{eq:ces2}\\
c_{\pi} & =-\frac{1}{k_{B}T}\left[\frac{1}{d}\mathcal{S}_{1}-\frac{\p}{m}\left(\chi_{1}+\frac{h}{m}\chi_{2}\right)\right],\nonumber 
\end{align}
and 
\begin{align}
v_{1} & =\hat{\Gamma}_{0}, & v_{2} & =1+\frac{c_{v}}{k_{B}}\hat{\Gamma}_{1}, & v_{3} & =\frac{1}{d}+\hat{\Gamma}_{0}+\hat{\Gamma}_{1},\\
w_{1} & =\hat{\Gamma}_{2}, & w_{2} & =\hat{\Gamma}_{2}-1, & w_{3} & =\frac{h}{k_{B}T}\hat{\Gamma}_{2}-1, & w_{4} & =-\frac{h}{k_{B}T}\hat{\Gamma}_{2}.
\end{align}
Notice that the frame dependence is contained in (\ref{eq:ces2}) while the freedom of choosing a representation is enclosed in the coefficients $v_{i}$ and $w_{i}$.\footnote{Using Eq.~(\ref{eses}), Kovtun's frame-invariant quantities defined in Eqs.~(A.28) and (A.29) of~\cite{aGjSoS2024b} are 
$f_i = -\left(\frac{1}{d}-\frac{k_{B}}{c_{v}}\right)^{-1}\zeta v_{i}$ and $\ell_{j}=-\kappa w_{j}/h$, and they are independent of $\chi_{i}$, as expected. Moreover, the frame- and representation-invariant quantities defined in that reference are $\zeta$, $\lambda_{1}=-(k_{B}T/h^{2})\kappa$, $\lambda_{2}=(e/h^{2})\kappa$, $\lambda_{3}=\kappa/h$, and are independent of both $\chi_{i}$ and $\hat{\Gamma}_{i}$.} 
Furthermore, from Eqs.~(\ref{eq:n1-1}), (\ref{eq:j-1}) and (\ref{eq:e1-1})-(\ref{eq:tau-1}), the quantities $\Phi_{S}$ and $\Phi_{V}^{\alpha}$ can be rewritten as
\begin{align}
\Phi_{S} & =v_{1}\frac{\dot{n}}{n}+v_{2}\frac{\dot{T}}{T}+v_{3}\theta=\mathcal{F}+\hat{\Gamma}_{0}\left(\frac{\dot{n}}{n}+\theta\right)+\hat{\Gamma}_{1}\left(\frac{c_{v}}{k_{B}}\frac{\dot{T}}{T}+\theta\right),
\label{eq:phiS}\\
\Phi_{V}^{\alpha} & =-w_{3}a^{\alpha}-w_{2}\frac{D^{\alpha}T}{T}-w_{1}\frac{D^{\alpha}n}{n}-w_{4}\frac{q}{h}E^{\alpha}=\mathcal{F}^{\alpha}-\frac{h\hat{\Gamma}_{2}}{k_{B}T}\left(a^{\alpha}+\frac{D^{\alpha}\p}{nh}-\frac{q}{h}E^{\alpha}\right),\label{eq:phiV}
\end{align}
with $\mathcal{F}:=\dot{T}/T+\theta/d$ and $\mathcal{F}_{\nu}:=a_{\nu}+D_{\nu}T/T$.
On shell, $\Phi_S$ and $\Phi_V^\alpha$ are just given by $\mathcal{F}$ and $\mathcal{F}^\alpha$. Thus,
starting from the expression that arises by implementing the projection
method in the Chapman-Enskog expansion, we exhibited that the core
structure of the constitutive equations for a single species charged
gas (in the presence of a weak background electromagnetic field) relates
the scalar dissipative quantities with the combination $\mathcal{F}$,
the vector ones with $\mathcal{F}_{\nu}$, while the tensor one is
only coupled with $\sigma_{\mu\nu}$. The coefficients in such relations
are here written in terms of $\mathcal{S}_{1,3,5}$ which depend on
the microscopic model for the collision term in Boltzmann's equation.
This core structure, which depends solely on $\mathcal{S}_{1,3,5}$
and $\mathcal{F}$, $\mathcal{F}_{\nu}$, and $\sigma^{\mu\nu}$ arises
naturally in the trace-fixed particle frame for which the compatibility
conditions are given by $g_{1}=1$, $g_{2}=\gamma$, and $g_{3}=1$,
which implies $\chi_{1}=\chi_{2}=\chi_{3}=0$. Starting with this
particular structure, one can extend the results to an arbitrary frame
by choosing different values for $g_{i}$ which adds a contribution
from the homogeneous equation. As we have shown in this work, this
leads to a change in the coefficients which neither breaks the combinations
given by $\mathcal{F}$ and $\mathcal{F}^{\alpha}$ nor introduces
new derivative terms in the constitutive equations. As examples, we
have included the particular cases of the particle and energy frames,
addressing within them Eckart and Landau's frames, respectively. One
can then conclude that, from the kinetic theory point of view, a change
of frame can only lead to shifts in the coefficients that keep the
aforementioned combinations invariant.

However, in order to obtain the most general constitutive equations from kinetic theory, one needs to consider an additional freedom, which we referred to as the choice of representation. This entails allowing for nonzero values of the parameters $\hat{\Gamma}_{i}$ which multiply second-order terms in Knudsen's parameter. This leads to the final form of the constitutive equations~(\ref{eq:n1-1}), (\ref{eq:j-1}), and (\ref{eq:e1-1})-(\ref{eq:tau-1}). These relations coincide in structure with the ones proposed and thoroughly discussed in recent works within the context of the BDNK theories. However, we wish to emphasize that the derivation described in the present work is different form Refs.~\cite{Rocha2022,RDNR2024} and arises from projecting the right-hand side of the linearized Boltzmann equation onto the orthogonal complement of $\ker\mathcal{L}$ following Ref.~\cite{LaurePaper} and formally established in the relativistic scenario in Ref.~\cite{Hydrolimit}. Moreover, the method carried out here clearly separates the effects of changes of frames and representations and sheds light on the microscopic origin of such freedoms. As has been shown in previous works, the freedom of choosing the parameters $\hat{\Gamma}_{i}$ is fundamental for obtaining a well-posed Cauchy problem for which global equilibrium states are stable. In the context of the TFP frame this has been established in Refs.~\cite{aGjSoS2024a,aGjSoS2024b} for a large class of models which includes a simple gas with hard spheres' cross sections. For conditions leading to hyperbolic and causal evolution equations in unspecified frames, see for instance Refs.~\cite{pKovtun19,Bemfica2022,fSfA26}.

As a final comment, we clarify the relation of our formalism with thermodynamics by analyzing the entropy production term, 
\begin{align}
T\nabla_{\alpha}S_{IS}^{\alpha} & =-\left(ne^{(1)}-n^{(1)}e\right)\frac{\dot{T}}{T}-\p^{(1)}\theta-k_{B}Tn^{(1)}\frac{\dot{n}}{n}\nonumber \\
 & -\left(\mathcal{Q}^{\alpha}-h\mathcal{J}^{\alpha}\right)\left(a_{\alpha}+\frac{D_{\alpha}T}{T}\right)-h\mathcal{J}^{\alpha}\left(a_{\alpha}+\frac{D_{\alpha}\mathrm{p}}{nh}-\frac{q}{h}E_{\alpha}\right)-\mathcal{T}^{\alpha\beta}\sigma_{\alpha\beta},
\end{align}
where $S_{IS}^{\mu}$ is the Israel-Stewart entropy flux (shown in~\cite{Hydrolimit}
to coincide with the Boltzmann entropy flux up and including first-order
terms). Substituting the general constitutive relations~(\ref{Eq:GeneralCR1},\ref{Eq:GeneralCR2}),
one obtains after some calculations, 
\begin{align}
T\nabla_{\alpha}S_{IS}^{\alpha} & =\left\{ \mathcal{S}_{1}\mathcal{F}-\frac{p\chi_{1}}{m}\left(\frac{\dot{n}}{n}+\theta\right)-\frac{\p\chi_{2}}{m^{2}}\left[e\left(\frac{\dot{n}}{n}+\theta\right)+k_{B}T\left(\frac{c_{v}}{k_{B}}\frac{\dot{T}}{T}+\theta\right)\right]\right\} \frac{\Phi_{S}}{k_{B}T}\nonumber \\
 & +\left\{ \frac{\mathcal{S}_{5}}{4}\mathcal{F}_{\alpha}-\frac{d\p h\chi_{3}}{2m^{2}}\left(a_{\alpha}+\frac{D_{\alpha}\mathrm{p}}{nh}-\frac{q}{h}E_{\alpha}\right)\right\} \frac{\Phi_{V}^{\alpha}}{k_{B}T}+\frac{\mathcal{S}_{3}}{k_{B}T}\sigma^{\alpha\beta}\sigma_{\alpha\beta},\label{eq:1}
\end{align}
where $\Phi_{S}$ and $\Phi_{V}^{\alpha}$ are given in Eqs.~(\ref{eq:phiS}) and (\ref{eq:phiV}).
In view of the balance equations and Eq.~(\ref{eses}), this yields
\begin{equation}
\nabla_{\alpha}S_{IS}^{\alpha}=\frac{\zeta}{T}\frac{\mathcal{F}^{2}}{\left(\frac{1}{d}-\frac{k_{B}}{c_{v}}\right)^{2}}+\frac{\kappa}{T}\mathcal{F}^{\alpha}\mathcal{F}_{\alpha}+\frac{2\eta}{T}\sigma^{\alpha\beta}\sigma_{\alpha\beta}+\mathcal{O}(\partial^{3}),
\end{equation}
where $\mathcal{O}(\partial^{3})$ refers to third-order off-equilibrium terms. It is important to stress that this result is valid in any frame and representation and that the second-order terms are positive-definite. Rewriting $\mathcal{F}=\left(\frac{1}{d}-\frac{k_{B}}{c_{v}}\right)\theta+\frac{k_{B}}{c_{v}}\theta+\frac{\dot{T}}{T}$ and using the balance equations one obtains the usual relation in the Eckart theory.

\acknowledgments

We thank J. Félix Salazar and C. Gabarrete for fruitful comments and discussions. This work was supported by SECIHTI under Grant No. CBF-2025-G-1626. O.S. also acknowledges support from CIC grant No.~18315 to Universidad Michoacana de San Nicolás de Hidalgo.

\appendix

\section{General form of the coefficient $\mathcal{R}_{\alpha}\,^{\mu\nu}\left[\varUpsilon\right]$}
\label{App:RCoeff} 

In this appendix we justify the particular form~(\ref{Eq:RReduced-1})
of the coefficient $\mathcal{R}_{\alpha}\,^{\mu\nu}\left[\varUpsilon\right]=\left\langle \varUpsilon\left(\gamma\right)p_{\alpha},\mathcal{L}^{-1}\left[\left(p^{\mu}p^{\nu}\right){}^{\perp}\right]\right\rangle $
defined in Eq.~(\ref{Eq:RDef-1}). For this, we first recall the
definition~(\ref{LinearizedCO}) of the linearized collision operator
$\mathcal{L}$, and for notational simplicity we omit the dependency
on $x$ in what follows. Using Lorentz invariance (and the fact that
$\mathcal{F}$ and $d\sigma/d\Omega$ are Lorentz scalars) it is simple
to verify that the operator $T_{\Lambda}$, which transforms $\phi(p)$
to $T_{\Lambda}\phi(p):=\phi(\Lambda^{-1}p)$, satisfies 
\begin{equation}
\left(T_{\Lambda}\mathcal{L}\left[\phi\right]\right)(p_{1})=\int\limits_{P_{x}^{+}(m)}\int\limits_{S^{d-1}}\mathcal{F}\frac{d\sigma}{d\Omega}T_{\Lambda}f^{(0)}(p_{2})\left[T_{\Lambda}\phi(p_{1})+T_{\Lambda}\phi(p_{2})-T_{\Lambda}\phi(p_{1}^{*})-T_{\Lambda}\phi(p_{2}^{*})\right]d\Omega(\hat{q}^{*})\dvol_{x}(p_{2}),\label{LinearizedCOBis}
\end{equation}
for all Lorentz transformations $\Lambda$. Since $T_{\Lambda}f^{(0)}=f^{(0)}$
if $\Lambda$ keeps the velocity vector $u^{\mu}$ fixed, it follows
that $T_{\Lambda}$ commutes with $\mathcal{L}$, such that 
\begin{equation}
T_{\Lambda}\mathcal{L}[\phi]=\mathcal{L}[T_{\Lambda}\phi],
\end{equation}
for all $\Lambda\in G_{u}$, where $G_{u}$ denotes the orthogonal
subgroup of Lorentz transformations which keep $u^{\mu}$ fixed. Since
$T_{\Lambda}$ leaves $\ker\mathcal{L}$ invariant, it follows that
$T_{\Lambda}$ also commutes with $\mathcal{L}^{-1}$ for all $\Lambda\in G_{u}$.

We first apply these properties to a coefficient of the form 
\begin{equation}
\mathcal{Q}^{\mu\nu}=\left\langle \varUpsilon\left(\gamma\right),\mathcal{L}^{-1}\left[\left(p^{\mu}p^{\nu}\right){}^{\perp}\right]\right\rangle ,
\end{equation}
where $\varUpsilon$ is a given function depending only on $\gamma=-u^{\alpha}p_{\alpha}/m$.
Due to the linearity of $\mathcal{L}^{-1}$, it follows immediately
that $\mathcal{Q}^{\mu\nu}[\varUpsilon]$ transforms like a Lorentz
tensor. Moreover, using the fact that $T_{\Lambda}$ is unitary with
respect to the scalar product $\left\langle \cdot,\cdot\right\rangle $
and that $\gamma$ is invariant with respect to $T_{\Lambda}$ for
all $\Lambda\in G_{u}$, it follows that 
\begin{equation}
\mathcal{Q}^{\mu\nu}=\Lambda^{\mu}{}_{\alpha}\Lambda^{\nu}{}_{\beta}\mathcal{Q}^{\alpha\beta},\label{Eq:QInv}
\end{equation}
for all $\Lambda\in G_{u}$, that is, $\mathcal{Q}^{\mu\nu}$ is actually
invariant with respect to $G_{u}$. Expanding 
\begin{equation}
\mathcal{Q}^{\mu\nu}=Q_{0}u^{\mu}u^{\nu}+Q_{1}\Delta^{\mu\nu}+2u^{(\mu}Q_{2}^{\nu)}+Q_{3}^{\mu\nu},
\end{equation}
where $Q_{2}^{\mu}$ and $Q_{3}^{\mu\nu}$ are orthogonal to $u^{\mu}$
and $Q_{3}^{\mu\nu}$ is symmetric and trace-free, we see that the
first two terms in this expansion are already invariant with respect
to $G_{u}$, whereas it follows from Eq.~(\ref{Eq:QInv}) that $Q_{2}^{\mu}=\Lambda^{\mu}{}_{\alpha}Q_{2}^{\alpha}$
and $Q_{3}^{\mu\nu}=\Lambda^{\mu}{}_{\alpha}\Lambda^{\nu}{}_{\beta}Q_{3}^{\alpha\beta}$
for all $\Lambda\in G_{u}$, which in turn implies that $Q_{2}^{\mu}=0$
and $Q_{3}^{\mu\nu}=0$. Since $\mathcal{Q}^{\mu\nu}$ is trace-free
by definition, it follows that 
\begin{equation}
\mathcal{Q}^{\mu\nu}=Q_{0}\left(u^{\mu}u^{\nu}+\frac{1}{d}\Delta^{\mu\nu}\right).
\end{equation}
Applying a similar argument to the contractions of $\mathcal{R}_{\alpha}\,^{\mu\nu}\left[\varUpsilon\right]$
with $u^{\alpha}$ and $\Delta^{\alpha\beta}$, one finds that 
\begin{equation}
\mathcal{R}_{\alpha}\,^{\mu\nu}\left[\varUpsilon\right]=R_{1}u_{\alpha}\left(u^{\mu}u^{\nu}+\frac{1}{d}\Delta^{\mu\nu}\right)+R_{2}\Delta_{\alpha}{}^{(\mu}u^{\nu)},
\end{equation}
which is of the desired form~(\ref{Eq:RReduced-1}). A similar argument applies to the decomposition of $\mathcal{S}^{\alpha\beta\mu\nu}$ in Eq.~(\ref{S}).

\bibliographystyle{unsrt}
\bibliography{refs_kinetic}

\end{document}